\documentclass[12pt]{revtex4}

\usepackage{mathbbol,mathrsfs,graphicx}
\begin{document}

\title[Spin and Orbital angular momentum propagation in anisotropic media: theory]{Spin and Orbital angular momentum propagation in anisotropic media: theory}

\author{Antonio Pic\'on$^{1,3*}$, Albert Benseny$^{1}$, Jordi Mompart$^{1}$, and Gabriel F. Calvo$^{2}$}

\address{$^{1}$Grup d'\`{O}ptica, Universitat Aut\`{o}noma de Barcelona, E-08193 Bellaterra (Barcelona), Spain}
\address{$^{2}$Departamento de Matem\'aticas, ETSI Caminos, Canales y Puertos \& IMACI-Instituto de Matem\'atica Aplicada a la Ciencia y la Ingenier\'{\i}a, Universidad de Castilla-La Mancha, E-13071 Ciudad Real, Spain}
\address{$^{3}$ (Actual address) JILA and Department of Physics, University of Colorado at Boulder, Boulder, Colorado 80309-0440, USA,* apicon@jilau1.colorado.edu}
\begin{abstract}
This paper is devoted to study the propagation of light beams carrying orbital angular momentum in optically anisotropic media. We first review some properties of homogeneous anisotropic media, and describe how the paraxial formalism is modified in order to proceed with a new approach dealing with a general setting of paraxial propagation along uniaxial inhomogeneous media. This approach is suitable for describing the space-variant-optical-axis phase plates. 
\end{abstract}

\maketitle

\section{Introduction}

In the past few years the research field of generation, manipulation and characterization of helical beams has attracted considerable interest. Under suitable generation conditions, these beams can transport angular momentum ({\em spin} and {\em orbital}) along their propagation direction \cite{allen_orbital_1992,calvo_quantum_2006}. These beams have triggered a wide range of applications (see the other papers in this special issue). Here, we examine the interplay between spin (SAM) and orbital angular momentum (OAM) of light upon propagation through anisotropic media, which is not only interesting from the fundamental viewpoint, but it also yields novel applications as we will expand in the following.

During paraxial propagation in isotropic homogeneous media, both spin and OAM are conserved quantities \cite{allen_orbital_1992,calvo_quantum_2006}, and thereby no exchange of angular momentum between these two is expected. However, in anisotropic media the beam during its propagation can exhibit a coupling between SAM and OAM, even a lossless angular momentum transfer between both of them \cite{calvo_spin-induced_2007}. A proof-of-principle demonstration of spin-controlled-changes in the OAM of circularly-polarized Gaussian beams in the visible domain, using patterned nematic liquid crystals, was experimentally achieved in 2006~\cite{marrucci_optical_2006}. The strong optical (uniaxial) anisotropic properties displayed by nematic liquid crystals were exploited to achieve a controllable space-dependent anisotropy of the medium. Moreover, one can reorient the optical axis of these media by applying external forces \cite{goldstein_polarized_2003}. Hence, one can construct optical elements exhibiting a space-variant-optical axis, in the sense that the optical axis varies at each point, thus allowing the medium to act on polarization differently for each position. A straightforward application of this type of media is the creation of non-scalar helical waves, based on spatially nonuniform polarization transformations, such as the attractive radial polarization or azimuthal polarization light beams (\cite{bomzon_radially_2002, Machavariani_Effect_2007,kawauchi_helical_beams_2008} and references therein). The other main application is found in the realm of quantum information and communication. Photons can carry information encoded in different degrees of freedom: polarization (SAM), spatial modes (OAM) and energy. The aim of quantum communication (QC) is to transfer and distribute quantum (entangled) states among distant sites (nodes) of networks where they are further processed \cite{duan_long-distance_2001,zoller_quantum_2005}. Devices allowing the unitary (lossless) information transfer from one degree of freedom to another degree of freedom in the same photon can really improve the robustness of the QC network \cite{deng_quantum_2007}. Anisotropic media are promising tools to transfer the entanglement encoded in the polarization of photons to their spatial mode profile and vice versa \cite{nagali_quantum_2009, nagali_polarization_2009}. Although previous works have achieved impressive results in this line, it is still an open question how to construct truly unitary transformations. Probably more complicated space-variant-optical-axis phase plates making use of the anisotropy along the radial axis will be necessary to achieve such unitary transformations, in the same line as it was anticipated in Ref. \cite{calvo_manipulation_2008}.

\par
So far, due to the interesting applications arisen by the space-variant-optical-axis phase plates, it is worthwhile briefly reviewing them. Space-variant-optical-axis phase plates have traditionally been fabricated by subwavelength metal stripes~\cite{bomzon_radially_2002}, and used as polarizers. In these, when the period of the grating is smaller than the wavelength of the incident light beam, only the zeroth order is a propagating order \cite{lopez_wave-plate_1998}, and the gratings behave as layers of a uniaxial crystal, whose optical axes are perpendicular and parallel to the grating grooves. This alternative way of creating birefringent components yields polarizing beam splitters and quarter-wave plates in the visible domain \cite{lopez_wave-plate_1998}. Exploiting this alternative approach of fabrication, the Hasman's group has been constructing space-variant-optical-axis phase plates to examine polarization manipulations on light. We shall not focus on these polarization manipulations in order not to deviate from our main track. For further details the reader is referred to Refs. \cite{bomzon_radially_2002,berry_adiabatic_1987,bomzon_pancharatnam--berry_2001,zhan_properties_2006}, where the connection between vectorial vortices, scalar vortices, Pancharatnam's phase and adiabatic phase is done. 
\par
The phase plates of Hasman's group were fabricated for light beams in the midinfrared spectral domain. However, in 2006 Marrucci et al. \cite{marrucci_optical_2006} developed space-variant-optical-axis phase plates in the visible domain to prove a unidirectional spin-to-orbital angular momentum switching. Spin-to-orbital angular momentum switching has also been observed without using space-variant phase plates, for example in the non-paraxial regime \cite{zhao_spin_2007,chen_electro_2008}, or a method to control the transfer of spin by an externally applied dc electric field in an optically active medium~\cite{chen_electrically_2006} was also proposed.
\par
Space-variant-optical-axis phase plates were firstly addressed under the Jones matrix formalism \cite{goldstein_polarized_2003,bomzon_pancharatnam--berry_2001}. In a previous work, we used a paraxial approach to prove a strong modulation in the spin and OAM changes when varying the traversed optical path \cite{calvo_spin-induced_2007}, but the development of such an approach was not presented. Here, we present for the first time an alternative approach based on the vectorial paraxial propagation of helical beams in space-variant-optical-axis media, which enables us to describe in a remarkable simple way the mechanism of spin-to-orbital angular momentum switching. In section \ref{sec:Paraxial_UAM_OA} we present an extension of the theory by Ciattioni et al. \cite{ciattoni_vectorial_2001,ciattoni_optical_2003} to treat the paraxial propagation in homogeneous uniaxial media. In section \ref{sec:Paraxial_IUAM_OA} we extend the previous formalism to deal with the paraxial propagation in inhomogeneous uniaxial media, which constitutes a very convenient way to describe the space-variant-optical-axis phase plates. Finally in section \ref{sec:Application} we conclude with a particular application of the derived formalism: the spin and OAM variation of Laguerre-Gaussian beams along the so-called $q$-plate.


\section{Paraxial propagation in uniaxial media} \label{sec:Paraxial_UAM_OA}

This section is devoted to the propagation of light beams carrying OAM in optically anisotropic media. We will address the interesting case of paraxial propagation along uniaxially homogeneous media, in an analogous way as the previous works of Ciattioni et al. \cite{ciattoni_vectorial_2001,ciattoni_optical_2003}, but here developing the general case where the optical axis can have any direction with respect to the wave propagation. This formalism will be extended to uniaxially inhomogeneous media, presenting simple and compact analytical formulas allowing to deal with a wide range of problems. The reader interested in the propagation of cylindrical partial waves along bianisotropic media is referred to \cite{novitsky2006}.
\par
In anisotropic media the Helmholtz wave equation satisfied by the electric field ${\bf E}={\bf E}({\bf r},t)$, when considering a monochromatic wave packet ${\bf E}({\bf r},t)={\bf E}({\bf r}) \, e^{-i\omega t}$, is:
\begin{eqnarray}\label{HelmholtzAM} 
{\nabla}^{2} {\bf E} - {\nabla}({\nabla} \cdot {\bf E}) + \frac{\omega^{2}}{c^{2}} \hat{\varepsilon} (\omega) {\bf E} = 0 \; ,
\end{eqnarray}
with $\omega$ being the frequency of the monochromatic wave, $c$ the speed of light, and $\hat{\varepsilon}$ the relative dielectric tensor. Notice that we explicitly include the possibility of a frequency dependence for $\hat{\varepsilon}$. In anisotropic media, the third term in Eq. (\ref{HelmholtzAM}) couples the field components, whereas the second term, which is often neglected in isotropic media, plays a relevant contribution here. When $\hat{\varepsilon}$ is a symmetric tensor, we can always choose a reference frame where the electrical permittivity is expressed in a diagonal form $\hat{\varepsilon}={\rm diag}(\varepsilon_{x},\varepsilon_{y},\varepsilon_{z})$. It is well known that two plane-wave families exist satisfying Eq. (\ref{HelmholtzAM}) for a given propagation direction, having different velocity phase and polarization. These two families are the so-called ordinary and extraordinary plane waves. Uniaxial media are materials with the cylindrical symmetry $\hat{\varepsilon}=\rm diag(\varepsilon_{\perp},\varepsilon_{\perp},\varepsilon_{\parallel})$ around the so-called optical axis (OA), corresponding to $\varepsilon_{\parallel}$, in this case the OA is in the $z$-direction. Defining the wave vector of a plane wave as ${\bf k}= \frac{\omega}{c} n \, {\bf u} = \frac{\omega}{c} \, {\bf n}= k_{0} \, {\bf n}$ (${\bf u}$ is a unitary vector pointing along the propagation direction, $n$ is the so-called refractive index, and ${\bf n}$ is the vector of refraction), Eq. (\ref{HelmholtzAM}) imposes two conditions in the plane wave vectors:
\begin{eqnarray}
        \label{UniaxialEq1}
        n^{2} = \varepsilon_{\perp} \; ,\\
        \label{UniaxialEq2}
        \frac{n_{x}^{2}+n_{y}^{2}}{\varepsilon_{\parallel}} +  \frac{n_{z}^{2}}{\varepsilon_{\perp}} = 1 \; .
\end{eqnarray}
In uniaxial media is common to define $\varepsilon_{\perp}\equiv n_{o}^{2}$ and $\varepsilon_{\parallel}\equiv n_{e}^{2}$, where $n_{o}$ and $n_{e}$ are the so-called ordinary and extraordinary refractive indexes. The dispersion relations (\ref{UniaxialEq1}) and (\ref{UniaxialEq2}) are the main equations to describe uniaxial anisotropic media, which yield two solutions related to the refractive index. The first solution, which satisfies Eq. (\ref{UniaxialEq1}), does not depend on the direction of the wave vector (ordinary waves), at variance with the solution satisfying Eq. (\ref{UniaxialEq2}), which depends on the wave propagation direction (extraordinary waves). Equation (\ref{UniaxialEq2}) is valid in the reference frame where the electrical permittivity is diagonal. In the general case,  the OA points along an arbitrary direction ${\bf u}_{\rm OA}=(\cos\varphi \sin\theta, \sin\varphi \sin\theta, \cos\theta)=\cos\varphi \sin\theta \,\hat{\bf i} + \sin\varphi \sin\theta\, \hat{\bf j}+ \cos\theta\, \hat{\bf k}$, with $\hat{\bf i}$, $\hat{\bf j}$, $\hat{\bf k}$ being unitary and orthogonal vectors associated with the laboratory frame axes.  The dispersion relation (\ref{UniaxialEq2}) is modified and can be easily obtained considering a rotation of the laboratory frame in order to fix the $z$-axis along the OA direction, so that one can resort to Eq. (\ref{UniaxialEq2}) with just a simple coordinates transformation. Here we have chosen the $\theta$ angle rotation around the $y$-direction (laboratory frame) and then a $\varphi$ angle rotation around the $z$-direction (laboratory frame), obtaining the general dispersion relation
\begin{eqnarray}  \nonumber
k_{0}^{2}  &=& \frac{k_{z}^{2}}{n_{e}^{2}(\theta)} + \frac{2k_{z}\cos\theta \sin\theta}{\Delta n^{2}} \left(k_{x} \cos\varphi + k_{y} \sin\varphi \right) \\
 &&+ \frac{(k_{x}\cos\varphi+k_{y}\sin\varphi)^{2}}{n_{e}'^{2}(\theta)} + \frac{(k_{x}\sin\varphi-k_{y}\cos\varphi)^{2}}{n_{e}^{2}}, \label{UniaxialEq2ForGeneralDirection}
\end{eqnarray}
where we have defined
\begin{eqnarray}
\hspace*{-10mm} \frac{1}{n_{e}^{2}(\theta)} \equiv \frac{\sin^{2}\theta}{n_{e}^{2}}+ \frac{\cos^{2}\theta}{n_{o}^{2}}\, , \quad \frac{1}{n_{e}'^{2}(\theta)} \equiv \frac{\cos^{2}\theta}{n_{e}^{2}}+ \frac{\sin^{2}\theta}{n_{o}^{2}}\, , \quad \frac{1}{\Delta n^{2}}
       \equiv  \left(\frac{1}{n_{o}^{2}}-\frac{1}{n_{e}^{2}} \right)\! . \label{def:Deltan,n'}
\end{eqnarray}
When $\theta = 0$ (the OA is parallel to the $z$-axis) one recovers Eq. (\ref{UniaxialEq2}). The two plane wave families can be expressed as 
\begin{eqnarray}
        \nonumber
        {\bf E}_{o}^{(\pm)}({\bf r}) = \tilde{\bf E}_{o} ({\bf q}) \exp [i \, {\bf q} \cdot {\bf r_{\perp}} \pm i \, (k_{0}^{2} n_{o}^{2} - q^{2})^{1/2} z]\, ,  \\
        {\bf E}_{e}^{(\pm)}({\bf r}) = \tilde{\bf E}_{e} ({\bf q}) \exp [i \, {\bf q} \cdot {\bf r_{\perp}} \pm i \, k_{z}(k_{x},k_{y}) z] \, , \label{PlanesWavesGeneral}
\end{eqnarray}
where ${\bf q} = k_{x} \, \hat{\bf i} + k_{y} \, \hat{\bf j}$ is the transverse wave vector, ${\bf r}_{\perp} \equiv x \, \hat{\bf i} + y \, \hat{\bf j} = (r \cos\phi, r\sin\phi)$ is the transverse coordinate, the function $k_{z}(k_{x},k_{y})$ is given by Eq. (\ref{UniaxialEq2ForGeneralDirection}), $\tilde{\bf E}_{o}({\bf q})$ and $\tilde{\bf E}_{e}({\bf q})$ are the ordinary and extraordinary polarization vectors (we will explain below how to calculate them). Then, a monochromatic plane wave is completely described by the transverse components $k_{x}$ and $k_{y}$. Any electric field propagating in an uniaxially anisotropic media, which is a solution of the Helmholtz Eq. (\ref{HelmholtzAM}), can be written as a superposition of ordinary and extraordinary plane waves (\ref{PlanesWavesGeneral}), 
\begin{eqnarray}\nonumber
        {\bf E}({\bf r}) = \int d^{2}{\bf q} &&\left\{\tilde{\bf E}_{o}({\bf q}) \exp [i \, {\bf q} \cdot {\bf r_{\perp}} \pm i \, (k_{0}^{2} n_{o}^{2} - q^{2})^{1/2} z] \right. \\
        &&+\left. \tilde{\bf E}_{e}({\bf q}) \exp [i \, {\bf q} \cdot {\bf r_{\perp}} \pm i \, k_{z}(k_{x},k_{y})z] \right\} . \label{ElectricFieldGC}
\end{eqnarray}
Hence, assuming that a (paraxial) beam is propagating along a well-defined direction, in this case $z$-direction, this can be perfectly decomposed as in Eq. (\ref{ElectricFieldGC}), but under the paraxial wave approximation ($k_{z} \gg k_{x},k_{y}$) we can rewrite Eq. (\ref{ElectricFieldGC}) in a simpler form that describes the propagation of any paraxial beam along the uniaxial medium. We start by performing the paraxial wave approximation on Eq. (\ref{UniaxialEq2ForGeneralDirection}) isolating $k_{z}(k_{x},k_{y})$. Therefore we obtain
\begin{eqnarray}  \label{kz_vs_kx,ky}
\hspace*{-5mm}k_{z}(k_{x},k_{y}) \rightarrow -q_{x} \cos\theta \sin\theta \; \frac{n_{e}^{2}(\theta)}{\Delta n^{2}} \pm k_{0}n_{e}(\theta)\!\left[ 1- \frac{n_{e}^{2}(\theta)}{2n_{e}^{2}n_{o}^{2}k_{0}^{2}} \, q_{x}^{2} - \frac{1}{2n_{e}^{2}k_{0}^{2}} \, q_{y}^{2} \right]\!  ,
\end{eqnarray}
where we have defined
\begin{eqnarray} \nonumber
q_{x} \equiv k_{x}\cos\varphi+k_{y}\sin\varphi \, , \\
 q_{y} \equiv k_{x}\sin\varphi-k_{y}\cos\varphi \, . \label{qxandqy}
\end{eqnarray}
\par
Now, instead of $k_{x}$ and $k_{y}$, we use henceforth $q_{x}$ and $q_{y}$, which depend of the OA-direction, in particular, on the $\varphi$ variable (which refers to rotations around the $z$ axis). 
\par
Secondly, we need to calculate the polarization vectors $\tilde{\bf E}_{o}({\bf q})$ and $\tilde{\bf E}_{e}({\bf q})$ in equation (\ref{ElectricFieldGC}). For the sake of brevity we will only sketch how to derive them. The ordinary polarization is perpendicular to the wave vector and the OA, thus $\tilde{\bf E}_{o} ({\bf q})  \propto {\bf k} \times {\bf u}_{OA}$.  On the other hand, the extraordinary polarization is perpendicular to the Poynting vector and the ordinary polarization, $\tilde{\bf E}_{e} ({\bf q})  \propto \tilde{\bf E}_{o} \times {\bf S}$. The direction of the Poynting vector is parallel to the gradient of the wave vector surface given by Eq. (\ref{UniaxialEq2ForGeneralDirection}). Knowing the direction of the two polarizations, we can finally obtain $\tilde{\bf E}_{o}({\bf q})$ and $\tilde{\bf E}_{e}({\bf q})$ by imposing the boundary conditions; the Fourier transform of the electric field (\ref{ElectricFieldGC}) at $z=0$ is
\begin{eqnarray}\label{FTGeneralSolutionofE}
\tilde{{\bf E}}({\bf q}) = \frac{1}{(2\pi)^{2}} \int d^{2} {\bf r}'_{\perp} \exp(-i\, {\bf q} \cdot {\bf r}'_{\perp}) \, {\bf E} ({\bf r'_{\perp}},0) \; ,
\end{eqnarray}
which is equal to $\tilde{{\bf E}}({\bf q})=\tilde{\bf E}_{o}({\bf q})+\tilde{\bf E}_{e}({\bf q})$. Here we provide the results within the paraxial approximation, distinguishing two well-defined regimes; when $\theta = 0$ and $\theta \ne 0$. If $\theta = 0$ (the OA is parallel to the $z$-axis), the polarization vectors reduce to
\begin{eqnarray}
        \label{PolarizationVectorsGDParaxial} \nonumber
        \tilde{\bf E}_{o} ({\bf q}) &=& \frac{1}{q^{2}} \left[(k_{y}^{2} \tilde{E}_{x} - k_{x}k_{y} \tilde{E}_{y}) \; \hat{{\bf i}}+
        (k_{x}^{2} \tilde{E}_{y} - k_{x}k_{y} \tilde{E}_{x})\; \hat{{\bf j}} \right]  ,\\ \label{PolarizationVectorsGDParaxial_1}
        \tilde{\bf E}_{e} ({\bf q}) &=& \frac{1}{q^{2}} \left[(k_{x}^{2} \tilde{E}_{x} + k_{x}k_{y} \tilde{E}_{y}) \; \hat{{\bf i}}+
        (k_{y}^{2} \tilde{E}_{y} + k_{x}k_{y} \tilde{E}_{x})\; \hat{{\bf j}} \right]  ,
\end{eqnarray}
which are in agreement with previous results obtained in Ref. \cite{ciattoni_vectorial_2001}. On the other hand, when $\theta \ne 0$, the polarization vectors become 
\begin{eqnarray} \nonumber
        \tilde{\bf E}_{o} ({\bf q}) &=& (\cos\varphi \, \tilde{E}_{y} -\sin\varphi \, \tilde{E}_{x}) \left[-\sin\varphi \; \hat{{\bf i}}+
        \cos\varphi \; \hat{{\bf j}} \right]  , \\ \label{PolarizationVectorsGDParaxial_2}
        \tilde{\bf E}_{e} ({\bf q}) &=& (\cos\varphi \, \tilde{E}_{x}+\sin\varphi \, \tilde{E}_{y})\left[\cos\varphi \; \hat{{\bf i}}+
        \sin\varphi \; \hat{{\bf j}}- \frac{\cos\theta \sin\theta}{\Delta n^{2}} \; \hat{{\bf k}}\right]  ,
\end{eqnarray}
with an explicit dependence on the OA-direction. In this case there is a $z$ component in the extraordinary part that cannot be neglected even in the paraxial approximation.
\par
Summarizing, we can expand any electric field into ordinary and extraordinary parts, as we did in Eqs. (\ref{ElectricFieldGC}). Hence, in the paraxial approximation (we still consider the paraxial wave propagating along $z$) the $z$ and $xy$ components of the electric field are 
\begin{eqnarray}
        \nonumber
        {\bf E}_{\perp o} ({\bf r_{\perp}},z) &=& e^{ik_{0}n_{o}z} \int d^{2}{\bf q} \;
        e^{i \, {\bf q} \cdot {\bf r_{\perp}} - i  \frac{q^{2}}{2k_{0}n_{o}}z} \, \hat{P}_{o} \tilde{{\bf E}}_{\perp}({\bf q}) \equiv e^{ik_{0}n_{o}z} {\bf A}_{\perp o} ({\bf r}_{\perp},z) \; ,  \\ \nonumber
        {\bf E}_{zo} ({\bf r_{\perp}},z) &=& 0 \; , \\ \nonumber
        {\bf E}_{\perp e} ({\bf r_{\perp}},z) &=& e^{ik_{0}n_{e}(\theta)z} \int d^{2}{\bf q} \;
        e^{i \, {\bf q} \cdot {\bf r_{\perp}} - i \, n_{e}(\theta) \left( \frac{q_{y}^{2}}{2n_{e}^{2}k_{0}}+ \frac{q_{x}^{2}}{2n_{e}^{2}k_{0}}
        \frac{n_{e}^{2}(\theta)}{n_{o}^{2}}\right)z} e^{-i \, q_{x} s\theta c\theta \;  \frac{n_{e}^{2}(\theta)}{\Delta n^{2}}z}  \\  \nonumber
        &\times& \hat{P}_{e} \tilde{{\bf E}}_{\perp}({\bf q}) \equiv e^{ik_{0}n_{e}(\theta) z} {\bf A}_{\perp e} ({\bf r}_{\perp},z)\; , \\ \nonumber
        {\bf E}_{ze} ({\bf r_{\perp}},z) &=& - \frac{\cos\theta \sin\theta}{\Delta n^{2}} \;
        e^{ik_{0}n_{e}(\theta)z} \int d^{2}{\bf q} \, (\cos\varphi \, \tilde{E}_{x}({\bf q})+\sin\varphi \, \tilde{E}_{y}({\bf q})) \\
        &\times&
        e^{i \, {\bf q} \cdot {\bf r_{\perp}} - i \, n_{e}(\theta) \left( \frac{q_{y}^{2}}{2n_{e}^{2}k_{0}}+ \frac{q_{x}^{2}}{2n_{e}^{2}k_{0}}
        \frac{n_{e}^{2}(\theta)}{n_{o}^{2}}\right)z} e^{-i \, q_{x} s\theta c\theta \;  \frac{n_{e}^{2}(\theta)}{\Delta n^{2}}z} \; ,    \label{GeneralEforANYdirection}
\end{eqnarray}
where the projector operators $\hat{P}_{o}$ and $\hat{P}_{e}$ are given below in Eqs. (\ref{ProjectorsGC}). Recall that the $q_{x}$ and $q_{y}$ variables, defined in Eq. (\ref{qxandqy}), depend on the OA direction. Moreover, we note again the presence of a $z$ component in the extraordinary field (\ref{GeneralEforANYdirection}) which cannot be neglected. However, in many systems, where the extraordinary refractive index does not differ much from the ordinary refractive index, the $1/\Delta n^{2}$ term is quite small (see Eq.~\ref{def:Deltan,n'}), and we can neglect the $z$ component of the electric field (when the OA is collinear or orthogonal to the wave propagation, it is a fairly good approximation). Let us examine more features. The ordinary part does not depend on the OA direction, as one would expect in uniaxial media. However, the extraordinary part has suffered a noticeable change. As the paraxial wave is composed by a plane wave times a slowly varying amplitude ${\bf A}_{\perp e} ({\bf r}_{\perp},z)$ (for ordinary waves this is  ${\bf A}_{\perp o} ({\bf r}_{\perp},z)$), we expect that its plane wave-like part will depend on the propagation direction, with a refractive index given by $n_{e}(\theta)$ (for ordinary waves $n_{o}$). The second main remark is related to the slowly varying amplitude ${\bf A}_{\perp e} ({\bf r}_{\perp},z)$, which does not obey the standard (isotropic-like) paraxial wave equation, due to the lack of symmetry in the transverse plane during the propagation. The slowly varying amplitudes obey
\begin{eqnarray} \nonumber
\left( i \frac{\partial}{\partial z} + \frac{1}{2k_{0}n_{o}} \nabla_{\perp}^{2} \right) {\bf A}_{\perp o} = 0 \; , \\
        \label{ParaxialEquationAM_General}
\left[ i \frac{\partial}{\partial z} + \frac{n_{e}^{3}(\theta)}{2k_{0}n_{o}^{2}n_{e}^{2}} \partial_{r_{x}}^{2}+ \frac{n_{e}(\theta)}{2k_{0}n_{e}^{2}} \partial_{r_{y}}^{2}- i \sin\theta \cos\theta \;  \frac{n_{e}^{2}(\theta)}{\Delta n^{2}} \partial_{r_{x}} \right] {\bf A}_{\perp e} = 0 \; ,
\end{eqnarray}
where we have defined,
\begin{eqnarray} \nonumber
r_{x} &\equiv& x\cos\varphi + y \sin\varphi \; ,\\ 
r_{y} &\equiv& x\sin\varphi - y \cos\varphi \; . \label{rxandry}
\end{eqnarray}
The rotated variables of Eq. (\ref{rxandry}) simply come from the rotated momenta variables defined in Eq. (\ref{qxandqy}). 
Another interesting point is the occurrence of a new term in the phase of the extraordinary wave, proportional to $q_{x}$ (originated by the walk-off effect). This term is responsible for a transversal translation of the extraordinary wave during propagation. But most surprising is the fact that projectors $\hat{P}_{o}$ and $\hat{P}_{e}$ are 
\begin{eqnarray}
\nonumber \hat{P}_{o}=\frac{1}{q^{2}} \left[
\begin{array}{cc}
k_{y}^{2}  &  -k_{x}k_{y} \\
-k_{x}k_{y}  &  k_{x}^{2} \\
\end{array}
 \right] , 
 \\
\hat{P}_{e}=\frac{1}{q^{2}} \left[ \begin{array}{cc}
k_{x}^{2}  &  k_{x}k_{y} \\
k_{x}k_{y}  &  k_{y}^{2} \\
\end{array} \right] \; , \label{Projector_theta_0}
\end{eqnarray}
when $\theta = 0$, but in the case when $\theta \neq 0$, projectors collapse to
\begin{eqnarray}
\nonumber \hat{P}_{o}= \left[
\begin{array}{cc}
\sin^{2}\varphi  &  -\sin\varphi \cos\varphi \\
-\sin\varphi \cos\varphi  &  \cos^{2}\varphi \\
\end{array}
 \right] = \frac{\hat{\mathbb{1}} - \hat{\mathcal{R}}_{z}(\varphi)
 \hat{\sigma}_{z} \hat{\mathcal{R}}_{z}(-\varphi)}{2} \; ,
  \\ \label{ProjectorsGC}
\hat{P}_{e}= \left[ \begin{array}{cc}
\cos^{2}\varphi &  \sin\varphi \cos\varphi \\
\sin\varphi \cos\varphi  &  \sin^{2}\varphi \\
\end{array}
 \right]= \frac{\hat{\mathbb{1}} + \hat{\mathcal{R}}_{z}(\varphi)
 \hat{\sigma}_{z} \hat{\mathcal{R}}_{z}(-\varphi)}{2} \; , \hspace{0,5cm}
\end{eqnarray}
where $\hat{\mathcal{R}}_{z}(\varphi)$ represents a $\varphi$ angle rotation around the $z$-axis and $\hat{\sigma}_{z}$ the usual third component Pauli matrix
\begin{eqnarray}
 \nonumber \hat{\mathcal{R}}_{z}(\varphi)  = \left[
\begin{array}{cc}
\cos\varphi  &  -\sin\varphi  \\
\sin\varphi  &  \cos\varphi  \\
\end{array}
 \right]\!  ,
  \\ \nonumber
\hat{\sigma}_{z} = \left[ \begin{array}{cc}
1 &  0 \\
0 &  -1\\
\end{array}
 \right]\!  ,\\ \label{Matrices}
\hat{R} \equiv \hat{\mathcal{R}}_{z}(\varphi)
 \hat{\sigma}_{z} \hat{\mathcal{R}}_{z}(-\varphi) = \left[
\begin{array}{cc}
\cos(2\varphi)  &  \sin(2\varphi)  \\
\sin(2\varphi)  &  -\cos(2\varphi)  \\
\end{array} \right] \! .
\end{eqnarray}
Notice that projectors (\ref{ProjectorsGC}) are independent of the $\theta$ variable.


\section{Paraxial propagation in inhomogeneous uniaxial media} \label{sec:Paraxial_IUAM_OA}

In section \ref{sec:Paraxial_UAM_OA} we have examined the propagation of a paraxial wave along anisotropic media, considering the OA constant in all the medium, in other words, the medium was homogenous and the OA has the same direction at any position of the medium. However, what happens if the medium is inhomogenous (i.e., the OA  depends on the position)? We will solve this problem under certain conditions. 
\par
Before addressing the inhomogenous problem, it is convenient to rewrite Eqs. (\ref{GeneralEforANYdirection}) in a suitable form, which will be essential for further calculations. Therefore, recalling the Fourier transform (\ref{FTGeneralSolutionofE}), we can substitute it into the transverse electric part of Eqs. (\ref{GeneralEforANYdirection}), obtaining 
\begin{eqnarray} \nonumber
{\bf E}_{\perp o} ({\bf r_{\perp}},z) &=& \frac{k_{0}n_{o}}{2 \pi i z} e^{i k_{0}n_{o}z}\! \int\! d^{2}{\bf r}'_{\perp} \hat{P}_{o} {\bf E}_{\perp}({\bf r}'_{\perp})\, e^{i \frac{k_{0}n_{o}}{2z}({\bf r}_{\perp} - {\bf r}_{\perp}')^{2}} \, ,        \\ \nonumber
{\bf E}_{\perp e} ({\bf r_{\perp}},z) &=& \frac{k_{0}n_{o}}{2 \pi i z} \frac{n_{e}^{2}}{n_{e}^{2}(\theta)} e^{i k_{0}n_{e}(\theta)z} \! \int\! d^{2}{\bf r}'_{\perp} \hat{P}_{e} {\bf E}_{\perp}({\bf r}'_{\perp}) \\
&\times& e^{i \frac{k_{0}}{2z} \frac{n_{e}^{2}}{n_{e}(\theta)} \left[  \frac{n_{o}^{2}}{n_{e}^{2}(\theta)} \left\{(r_{x}-r_{x}')- \sin\theta \cos\theta \frac{n_{e}^{2}(\theta)}{\Delta n^{2}} z \right\}^{2} + (r_{y}-r_{y}')^{2} \right]} \, ,  \label{FresnelPropagation}
\end{eqnarray}
where we have used the definitions of the rotated position variables (\ref{rxandry}). Notice the similitude of Eqs. (\ref{FresnelPropagation}) with the Fresnel integral in the isotropic case. The Fresnel integral describes the evolution of an input paraxial wave, taking into account its initial boundary conditions. It is basically composed by the product of the input paraxial wave with the Fresnel kernel (propagator). In anisotropic media, integrals (\ref{FresnelPropagation}), analogously to the Fresnel integral, describe the paraxial wave evolution. Hence, performing the integration with respect to the variable ${\bf r}'$, which corresponds to the transverse plane in $z=0$, we can determine the evolution of the wave for any $z$. First of all, projectors $\hat{P}_{o}$ and $\hat{P}_{e}$ select the ordinary and extraordinary part of the input electric field, depending on its polarization. After that, each part evolves independently with its corresponding ordinary and extraordinary kernel. Notice that the ordinary kernel is invariant under reference system rotations, as the usual Fresnel kernel. However, the extraordinary kernel depends on the OA direction, affecting thus the wave propagation. We can introduce the following new functions:
\begin{eqnarray} \nonumber
F_{o} &\equiv& e^{i k_{0}n_{o}z} e^{i \frac{k_{0}n_{o}}{2z} \vert {\bf r}_{\perp} - {\bf r}_{\perp}'\vert^{2}}\\ \label{functions_F}
F_{e} &\equiv& e^{i k_{0}n_{e}(\theta)z} e^{i \frac{k_{0}}{2z} \frac{n_{e}^{2}}{n_{e}(\theta)} \left[ \frac{n_{o}^{2}}{n_{e}^{2}(\theta)} \left\{(r_{x}-r_{x}')- \sin\theta \cos\theta \frac{n_{e}^{2}(\theta)}{\Delta n^{2}} z \right\}^{2} + (r_{y}-r_{y}')^{2} \right]}  ,
\end{eqnarray}
which, together with projectors (\ref{ProjectorsGC}) (we are not considering projectors (\ref{Projector_theta_0}), i.e. the case $\theta=0$, as the Fourier transform cannot be cast in the same analytical compact expression), allow us to simplify expressions (\ref{FresnelPropagation}) as
\begin{eqnarray} \nonumber
        {\bf E}_{\perp}({\bf r_{\perp}},z)  & = & {\bf E}_{\perp o} ({\bf r_{\perp}},z) + {\bf E}_{\perp e} ({\bf r_{\perp}},z) = \frac{k_{0}n_{o}}{2 \pi i z}\!
        \int\! d^{2}{\bf r}'_{\perp} \left\{\left[\frac{n_{e}^{2}}{n_{e}^{2}(\theta)}F_{e}+F_{o} \right]\frac{\hat{\mathbb{1}}}{2} \right. \\
       & + & \left. \left[\frac{n_{e}^{2}}{n_{e}^{2}(\theta)}F_{e}-F_{o} \right]\frac{\hat{\mathcal{R}}_{z}(\varphi)
 \hat{\sigma}_{z} \hat{\mathcal{R}}_{z}(-\varphi)}{2} \right\}{\bf E}_{\perp}({\bf r}'_{\perp}) \, . \label{FresnelIntegralsAM}
\end{eqnarray}
Equation (\ref{FresnelIntegralsAM}) has two main terms. The first one is a factor $\left[n_{e}^{2}/n_{e}^{2}(\theta) F_{e} + F_{o} \right]$ times the identity matrix, which does not change the input polarization. However, it couples both extraordinary and ordinary parts. On the other hand, the second term, $\left[n_{e}^{2}/n_{e}^{2}(\theta) F_{e} - F_{o} \right]$ times a rotation matrix, also combines the ordinary and the extraordinary part but, at the same time, it varies the input polarization. In the isotropic limit, when $n_{e}$ tends to $n_{o}$, we find that from $\left[n_{e}^{2}/n_{e}^{2}(\theta) F_{e} + F_{o} \right]$ one recovers the standard Fresnel kernel, whereas $\left[n_{e}^{2}/n_{e}^{2}(\theta) F_{e} - F_{o} \right]$ vanishes.
\par
Equation (\ref{FresnelIntegralsAM}) still describes homogenous anisotropic media, with an OA direction given by ($\theta$, $\varphi$). However, in an inhomogenous medium, the OA direction can depend on the position. Thus, both $\theta = \theta ({\bf r}')$ and $\varphi = \varphi ({\bf r}')$ are functions of the position. Inserting this dependence into Eq. (\ref{FresnelIntegralsAM}), and remembering that the function $F_{e}$ also depends on the OA direction, we can calculate the evolution of a paraxial wave in uniaxial anisotropic inhomogenous media. Of course, we consider that $\theta({\bf r}')$ and $\varphi ({\bf r}')$ are sufficiently smooth functions.
\par
The approach developed here is quite general and allows one to deal with paraxial beams propagating along inhomogeneous anisotropic media by using the integral propagator (\ref{FresnelIntegralsAM}). In particular, the integral propagator (\ref{FresnelIntegralsAM}) can be used for beams with a certain spatial structure carrying OAM. One can easily apply this integral propagator to describe the spin-to-orbital angular momentum switching observed experimentally in Ref. \cite{marrucci_optical_2006} by using space-variant-optical-axis phase plates. For the sake of completeness, we will address this problem with our formalism in the following section.


\section{Application: Propagation of a Laguerre-Gaussian Mode along a $q$-Plate} \label{sec:Application}

\par
In this section we consider a specific example, the propagation of a Laguerre--Gaussian (LG) beam through a space-variant-optical axis medium, whose OA is always perpendicular to the propagation direction (this means $\theta({\bf r}') = \pi/2$, see the previous section). Here we will focus on the, so-called, $q$-plates, which have the following linear OA dependence $\varphi ({\bf r}')=q\,\phi'+\alpha_{0}$ ($q$ and $\alpha_{0}$ are constants).
\par
Hence, the initial electric field in Eq. (\ref{FresnelIntegralsAM}) is written as ${\bf E}_{\perp}(r',\phi',z=0)$ $= (a\,\hat{\bf u}_{+}+ b\,\hat{\bf u}_{-})$ $LG_{\ell,p}(r',\phi')$, where $\hat{\bf u}_{\sigma}\equiv (\hat{\bf i}+i\sigma \hat{\bf j})/\sqrt{2}$ are the circular polarization vectors, and $\sigma=\pm1$ for right- and left-hand circularly polarization, respectively. The functions $LG_{\ell,p}(r',\phi')$ denote the LG modes~\cite{allen_orbital_1992,calvo_quantum_2006}, which comprisse two terms $LG_{\ell,p}(r',\phi')\equiv e^{i\ell \phi'} R_{\ell,p} (r')$, where $R_{\ell,p} (r')$ is a purely radial function depending on the two indices: $\ell$ and $p$ which account for the orbital angular momentum (topological charge) and the radial node number of the spatial mode, respectively~\cite{allen_orbital_1992,calvo_quantum_2006}. Using Eq. (\ref{FresnelIntegralsAM}) we can calculate the propagation of the Laguerre-Gaussian beam along the $q$-plate. The electric field can be separated into two parts ${\bf E}_{\perp}({\bf r_{\perp}},z)={\bf E}_{\perp}^{(1)}({\bf r_{\perp}},z)+{\bf E}_{\perp}^{(2)}({\bf r_{\perp}},z)^{(2)}$, where ${\bf E}_{\perp}^{(1)}({\bf r_{\perp}},z)$ preserves the initial polarization while that of ${\bf E}_{\perp}^{(2)}({\bf r_{\perp}},z)$ changes:
\begin{eqnarray}
        {\bf E}_{\perp}^{(1)}({\bf r_{\perp}},z)\!=\!\frac{k_{0}n_{o}}{4 \pi i z}\!
        \int\! d^{2}{\bf r}'_{\perp}  (F_{e}\!+\!F_{o})(a\,\hat{\bf u}_{+}+ b\,\hat{\bf u}_{-}) e^{i\ell \phi'}\!R_{\ell,p} (r') \; , \label{FresnelIntegralsAM_1} \\
        {\bf E}_{\perp}^{(2)}({\bf r_{\perp}},z)\!=\!\frac{k_{0}n_{o}}{4 \pi i z}\!
        \int\! d^{2}{\bf r}'_{\perp}  (F_{e}\!-\!F_{o})(b \,e^{-2i\varphi} \,\hat{\bf u}_{+}+ a\,e^{2i\varphi}\,\hat{\bf u}_{-}) e^{i\ell \phi'}\!R_{\ell,p} (r')\; . \label{FresnelIntegralsAM_2}
\end{eqnarray}
Note that the second term that changes the polarization is also introducing exponential terms that include the OA dependence  $\varphi (\phi',\alpha_{0})=q\,\phi'+\alpha_{0}$. In this case propagators $F_{o}$ and $F_{e}$ (\ref{functions_F}) can be expressed as
\begin{eqnarray} \nonumber
F_{o} &\equiv& e^{i k_{0}n_{o}z} e^{i \frac{k_{0}n_{o}}{2z} \vert {\bf r}_{\perp} - {\bf r}_{\perp}'\vert^{2}}\\ \label{functions_F}
F_{e} &\equiv& e^{i k_{0}n_{e}z} e^{i \frac{k_{0}n_{e} }{2z} \left[ \frac{n_{o}^{2}}{n_{e}^{2}} (r_{x}-r_{x}')^{2} + (r_{y}-r_{y}')^{2} \right]} \approx e^{i k_{0}n_{e}z} e^{i \frac{k_{0}(n_{o}^{2}+n_{e}^{2})}{4zn_{e}} \vert {\bf r}_{\perp} - {\bf r}_{\perp}'\vert^{2}} ,
\end{eqnarray}
where the last approximation in the extraordinary propagator can be done as long as the birefringence is not too large, that is $\vert n_{o}^{2}-n_{e}^{2} \vert \ll n_{o}^{2}+n_{e}^{2}$. The neglected part accounts for  a small astigmatism in the extraordinary part of the electric field.  We will focus on the angular part of the integrals (\ref{FresnelIntegralsAM_1}) and (\ref{FresnelIntegralsAM_2}) in order to analyze the spin and angular momentum of the beam along the $q$-plate. All the integrals in Eqs. (\ref{FresnelIntegralsAM_1}) and (\ref{FresnelIntegralsAM_2}) have the form
\begin{eqnarray} \label{Integral_Type_1}
\frac{k_{0}n_{o}}{4 \pi i z}e^{i k_{0}n_{1}z} \!\! \int_{0}^{\infty}\!\!\!\! r' dr' e^{i \frac{k_{0}n_{2}}{2z}(r^{2}+r'^{2})} R_{\ell,p} (r')
       \!\! \int_{0}^{2\pi} \!\!\!\! d\phi'  e^{-i \frac{k_{0}n_{2}}{z}(rr' \cos(\phi-\phi'))} e^{i\ell \phi'} \; ,
\end{eqnarray}
or
\begin{eqnarray}\label{Integral_Type_2}
\frac{k_{0}n_{o}}{4 \pi i z}e^{i k_{0}n_{1}z} \!\! \int_{0}^{\infty}\!\!\!\! r' dr' e^{i \frac{k_{0}n_{2}}{2z}(r^{2}+r'^{2})} R_{\ell,p} (r')
       \!\! \int_{0}^{2\pi} \!\!\!\! d\phi'  e^{-i \frac{k_{0}n_{2}}{z}(rr' \cos(\phi-\phi'))} e^{i(\ell \phi' \pm 2\varphi)} \; ,
\end{eqnarray}
where $n_{1}$ and $n_{2}$ are refractive indexes denoting $n_{o}$, $n_{e}$, or $(n_{0}^{2}+n_{e}^{2})/(2n_{e})$. By resorting to the Jacobi-Anger expansion we can easily deal with the angular factor
\begin{eqnarray} \label{Jacobi_Anger_E}
 e^{-i \frac{k_{0}n_{2}}{z}(rr' \cos(\phi-\phi'))} = \sum_{k=-\infty}^{\infty} (-i)^{k} J_{k}\left( \frac{k_{0} n_{2}}{z} r r' \right) e^{ik(\phi-\phi')} \; .
\end{eqnarray}
Therefore, by inserting the expansion (\ref{Jacobi_Anger_E}) in the integrals (\ref{Integral_Type_1}) and (\ref{Integral_Type_2}), and integrating with respect to the variable $\phi'$ (considering the $\varphi ({\bf r}')$ dependence in the $q$-plates), we can reduce the integrals to

\begin{eqnarray}
(-i)^{\ell+1} \; \frac{k_{0}n_{o}}{2 z}e^{i k_{0}n_{1}z}  e^{ i\ell \phi} \int_{0}^{\infty}\!\!\!\! r' dr' e^{i \frac{k_{0}n_{2}}{2z}(r^{2}+r'^{2})} R_{\ell,p} (r') J_{\ell}\left( \frac{k_{0} n_{2}}{z} r r' \right)  ,  \label{Integral_Type_1_r}
\end{eqnarray}
or
\begin{eqnarray}\nonumber
(-i)^{\ell\pm2q+1} \; \frac{k_{0}n_{o}}{2 z}e^{i k_{0}n_{1}z} e^{ \pm i2\alpha_{0}} e^{ i(\ell \pm 2q)\phi}\\ 
\times \int_{0}^{\infty}\!\!\!\! r' dr' e^{i \frac{k_{0}n_{2}}{2z}(r^{2}+r'^{2})} R_{\ell,p} (r') J_{\ell\pm2q}\left( \frac{k_{0} n_{2}}{z} r r' \right)  .  \label{Integral_Type_2_r}
\end{eqnarray}

By using integrals (\ref{Integral_Type_1_r}) and (\ref{Integral_Type_2_r}) we can analyze the propagation of the Laguerre-Gaussian beam along the $q$-plate given by Eqs. (\ref{FresnelIntegralsAM_1}) and (\ref{FresnelIntegralsAM_2}). The part of the propagation that does not modify the polarization of the initial beam is not changing the angular structure of the beam neither. Thus, the propagator (\ref{FresnelIntegralsAM_1}) is preserving the polarization and the angular structure, in other words, the spin and angular momentum of the paraxial beam remain intact. This explains why, in Ref. \cite{marrucci_optical_2006}, there was an experimentally observed part of the beam which exhibited unchanged polarization and angular structure. On the other hand, the propagator (\ref{FresnelIntegralsAM_2}) does indeed affect the polarization and alters the angular structure too. Therefore, a right- or left-handed circularly polarized LG beam changes its polarization to left- or right-handed circularly polarized respectively, and its topological charge from $\ell$ to $\ell + 2q$ in the case of left-handed circular polarization or to $\ell - 2q$ in the case of right-handed circular polarization. In other words, the variation of spin angular momentum and OAM is $-2\hbar$ and $2q\hbar$ respectively for an initial right-circularly polarized LG beam, resulting in a total variation of angular momentum of $2(q-1)\hbar$. For an initial left-circularly polarized LG beam the total variation of angular momentum is $-2(q-1)\hbar$, in complete agreement with the experiment performed in Ref. \cite{marrucci_optical_2006}. Moreover we can obtain the radial part of the LG beam propagating along the $q$-plate, which in general will be a superposition of LG modes with different radial number $p$ (the radial part of the LG mode is not preserved). Furthermore, this formalism is not only able to explain the spin-to-orbital angular momentum switching, one can also analytically calculate the angular momentum of light before and after the space-variant-optical-axis phase plates, as we first proved in Ref.~\cite{calvo_spin-induced_2007}.
\par
In this simple example we have showed how resourceful can be the developed formalism to address the paraxial propagation in inhomogeneous media.


\section{Conclusions}

To conclude, we have developed an approach, based on the vectorial paraxial propagation of helical beams, to deal with homogeneous and inhomogeneous anisotropic media. Within this approach, one can calculate the propagation of a paraxial wave through common anisotropic media, such as polarizers, and more complex optical elements such as space-variant phase plates which have shown promising applications~\cite{bomzon_radially_2002,marrucci_optical_2006,chen_electrically_2006,chen_electrically_2009}. This approach also allows to calculate the angular momentum of light, suitable for describing the OAM changes of the beam during propagation \cite{calvo_spin-induced_2007}. We have to remark that alternative approaches have been developed for some specific cases. Karimi et al.~\cite{karimi_light_2009} tried to avoid the approximation that $\theta({\bf r}')$ and $\varphi ({\bf r}')$ are very smooth functions in the case of an specific space-variant phase plate, the so-called $q$-plate. However, in order to do so, they neglected the term ${\nabla} \cdot {\bf E}$ in the anisotropic Helmholtz wave equation (\ref{HelmholtzAM}). This term is preserved in our approach. In a subsequent work, Vaveliuk~\cite{vaveliuk_nondiffracting_2009} showed that the term ${\nabla} \cdot {\bf E}$ cannot be neglected, providing a non-paraxial solution of such complicated anisotropic wave equation. Unfortunately such solutions are only valid for $q$-plates with $q=1$. 


\section*{Acknowledgements}

G.F.C. wishes to thank Junta de Castilla-La Mancha for financial support via Project PCI08-0093-6563.


\end{document}